\documentclass[11pt]{article}
\usepackage{moriond,epsfig}

\def\be{\begin{equation}}
\def\ee{\end{equation}}
\def\bea{\begin{eqnarray}}
\def\eea{\end{eqnarray}}

\begin{document}
\vspace*{4cm}
\title{Minimally Parametric Constraints on the Primordial Power Spectrum from Lyman-alpha}

\author{Simeon Bird}

\address{Institute of Astronomy and Kavli Institute for Cosmology, 
University of Cambridge, Cambridge, UK}

\maketitle
\abstracts{
Current analyses of the Lyman-alpha forest assume that the primordial power spectrum of density perturbations
obeys a simple power law, a strong theoretical assumption which should be tested.
Employing a large suite of numerical simulations which drop this assumption, we reconstruct the shape of the primordial power spectrum using Lyman-alpha data from the Sloan Digital Sky Survey (SDSS). Our method combines a minimally parametric framework with cross-validation, a technique used to avoid over-fitting the data.
Future work will involve predictions for the upcoming Baryon Oscillation Sky Survey (BOSS), 
which will provide new Lyman-alpha data with vastly decreased statistical errors. 
}

\section{Introduction}

The Lyman-$\alpha$ forest is the name given to a series of absorption lines in quasar spectra, caused by the 
scattering of photons via interaction with neutral hydrogen at redshifts $2-4$. At these redshifts, 
a large proportion of the baryon density of the universe is contained within hydrogen clouds. Most of the 
hydrogen is ionized, but a small fraction remains neutral, and absorbs photons via the Lyman-$\alpha$
transition.
Hence, the Lyman-$\alpha$ forest is sensitive to the matter power spectrum on scales from a few up to tens of Mpc, 
making it the only currently available probe of fluctuations at these weakly non-linear scales. 
A number of authors have examined the constraints obtainable from the Lyman-$\alpha$ forest in the past, 
including Croft et al \cite{croft}, Gnedin \& Hamilton \cite{gnedin}, 
Viel, Haehnelt \& Springel \cite{viel} .

Previous analyses of constraints from the Lyman-$\alpha$ forest have assumed that the primordial power spectrum is described by 
a nearly scale-invariant power law. 
This deserves further attention for a number of reasons. First, it is a strong assumption; if the 
data are inconsistent with it, derived constraints could be biased to some extent. 
Second, it is a generic prediction of inflationary models; hence, any test of a power law primordial power spectrum 
which cannot be attributed to data systematics is a test of inflation. Third, of all current datasets, 
the Lyman-$\alpha$ constrains the smallest cosmological scales; thus, it provides the best opportunity 
presently available to understand the overall shape of the power spectrum.
To do this, we shall reconstruct the primordial power spectrum in a minimally parametric way, 
using a technique called cross-validation 
to robustly recover the signal. If the data are in agreement with theoretical expectations, 
the recovered power spectrum will be nearly scale-invariant. In these Proceedings, we discuss
a minimally parametric framework for constraining the primordial matter power spectrum, 
the cross-validation technique, and the methodology for obtaining 
constraints from observations. Finally, some preliminary results are presented.

\section{Flux Power Spectrum}

In the case of Lyman-$\alpha$, the observable is not a direct measurement of the 
clustering properties of tracer objects, as in galaxy clustering, but the 
statistics of absorption along a number of quasar sightlines. 
It is easiest to work with the statistics of the flux, $\mathcal{F}$, defined as
\begin{equation}
        \mathcal{F} = \exp (-\tau),
        \label{eq:flux}
\end{equation}
where $\tau$ is the optical depth. The primary observable here is the one dimensional flux power spectrum, $P_\mathrm{F}$, 
\begin{eqnarray}
        P_\mathrm{F}(k) = | \tilde{\mathcal{F}}(k) |^2, 
\end{eqnarray}
where $\tilde{\mathcal{F}}$ is the Fourier transform of the flux, evaluated as a function of distance along 
the line of sight,
\begin{eqnarray}
        \tilde{\mathcal{F}}(k) = \int \mathcal{F}(x) e^{i kx} dx  \,.
        \label{eq:powerf}
\end{eqnarray}
Current constraints on $P_\mathrm{F}$ are given by McDonald et al \cite{McDonald}, determined from $\sim 3000$ SDSS quasar spectra.  

\begin{figure*}
\centering
\includegraphics[width=0.7\textwidth]{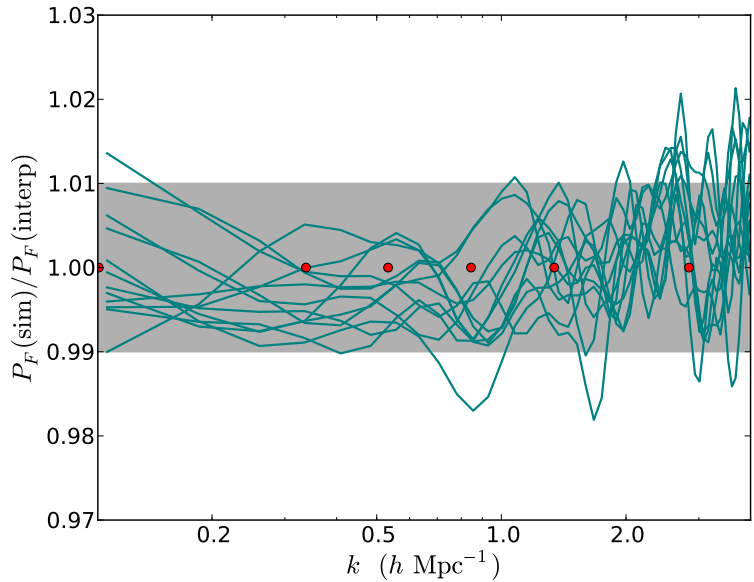}
\caption{The difference between the flux power spectrum as obtained from interpolation, using 
Eq. \ref{eq:fluxpower}, 
and directly by simulation. Each line represents simulation output at a different redshift bin, 
between $z=2.0$ and $z=4.2$. 
Red dots show the positions of spline knots. The grey band shows $1\%$ error bars.} 
\label{fig1}
\end{figure*}

In order to simulate the observable flux power spectrum from a given set of primordial fluctuations, 
a large $N$-body simulation is required. This 
makes it impractical to directly calculate $P_\mathrm{F}$ for every possible set of input parameters; 
instead simulations are run for a representative sample. Other results are obtained via interpolation, 
using the following scheme of Viel \& Haehnelt \cite{viel2}.
The flux power spectrum is assumed to be given by a Taylor expansion around some best-fit model.
For a vector of parameters $p_i$, with best-fit model parameters $p_i^0$, the flux 
power spectrum $P_\mathrm{F}$ is given by

\begin{equation}
        P_\mathrm{F}(p_i) = P_\mathrm{F}(p_i^0) + \Sigma_i (p_i-p_i^0)\frac{\partial P_\mathrm{F}}{\partial p_i} + \Sigma_i (p_i-p_i^0)^2 \frac{\partial^2 P_\mathrm{F}}{\partial p^2_i} \,.
        \label{eq:fluxpower}
\end{equation}
Numerical simulations are used to calculate the derivatives of the flux power spectrum.
Each parameter is varied independently, and the total change in the flux power spectrum is assumed to be 
a linear combination of the change due to each parameter, i.e.,

\begin{equation}
        \delta P_\mathrm{F} = \frac{\delta P_\mathrm{F}}{\delta p_1} \delta p_1 + \frac{\delta P_\mathrm{F}}{\delta p_2} \delta p_2 + \dots\,.
        \label{eq:variation}
\end{equation}
Figure \ref{fig1} shows the error due to this approximation for a sample input primordial power spectrum. 
The error is around $1\%$ on scales probed by current Lyman-$\alpha$ data ($k = 0.4 - 2\, h\, \mathrm{Mpc}^{-1}$), 
which is a small contribution to the total error, allowing us to proceed with confidence. Further checks on interpolation 
errors are in progress, and are expected to give similar results.

\section{Power Spectrum Reconstruction}

Previous analyses of the Lyman-$\alpha$ forest \cite{viel,McDonald2} have assumed 
the primordial power spectrum is a nearly scale-invariant power law, of the form

\begin{equation}
        P(k) = A_\mathrm{s}\left(\frac{k}{k_0}\right)^{n_\mathrm{s}-1}.
        \label{eq:pk}
\end{equation}
As discussed above, we seek to test whether the data supports this assumption by reconstructing 
the power spectrum with smoothing 
splines \footnote{Splines are piecewise cubic polynomials with globally continuous first and second derivatives, completely 
specified by their values at a series of knots, where the polynomials meet.}, 
as proposed in Sealfon et al \cite{sealfon}. Smoothing splines are used because they have good continuity 
properties and are particularly suited to formulation of a cross-validation penalty.

\section{Cross-Validation}

Any minimally parametric formalism, when applied to noisy data, runs the risk of over-fitting the data. 
One way to avoid this problem is a technique called cross-validation, described in Peiris \& Verde \cite{pv}. 
This technique assumes that noise in the data takes the form of additional small-scale structure, and thus 
power spectra with superfluous fluctuations should be penalised.
This penalty is implemented by adding an extra term to the likelihood function, $\mathcal{L}$;

\begin{equation}
        \log \mathcal{L} = \log \mathcal{L}(\mathrm{Data} | P(k)) + \lambda \int_k dk (P''(k))^2.
        \label{eq:penalty}
\end{equation}

Here $\lambda$, the penalty weight, is a free parameter. In the limit $\lambda \to \infty$ this likelihood 
becomes functionally identical to linear regression, while $\lambda \to 0$ is appropriate in the case of noiseless data.
In order to determine the optimal 
value for $\lambda$, the data points are first divided into two sets, the training set, or CV1, and the 
validation set, or CV2. CV1 and CV2 are composed of alternating data bins. 
Next, to calculate the CV score, a value is chosen for $\lambda$, and the best fit 
power spectrum based on the CV1 dataset is found. The $\chi^2$ is then calculated for this power spectrum 
with the CV2 dataset. This is repeated, replacing CV1 with CV2 and vice versa, and the CV score is the 
sum of both $\chi^2$ values.

The key to cross-validation is that signal in the CV1 dataset will predict signal in the CV2 dataset well, 
while noise in CV1 will predict noise in CV2 poorly. The optimal choice of $\lambda$ is therefore the one 
which allows maximal predictivity between CV1 and CV2; in other words, minimizes the CV score. 

\section{Results}\label{results}

We performed a large grid of $N$-body simulations using Gadget-II \cite{gadgetii} . Convergence checks were carried out to 
ensure $P_\mathrm{F}$ was not significantly affected by simulation settings \cite{heitmann} , such as particle 
resolution or box size. 
Initial conditions included a variety of input power spectra, on scales ranging from
$k=0.45 - 2 \,h\, \mathrm{Mpc}^{-1}$.

A significant departure from a power law primordial power 
spectrum translates to a detectable feature in the flux power spectrum, which is more noticeable at higher redshifts. 
This is due to the way in which the matter power spectrum evolves: a feature in the matter 
power spectrum will create extra non-linear growth on smaller scales, making the feature in 
$P_\mathrm{F}$ stand out less. The results of the simulations provide a mapping between primordial and flux power spectra,
which in turn provides a likelihood function for any given primordial power spectrum from SDSS data. The full data analysis, 
including cross-validation, is currently being carried out. 

\section{Future Prospects}

The best constraints on the flux power spectrum currently come from the Sloan Digital Sky Survey (SDSS
\cite{McDonald}), which contains $\sim 3000$ quasar sightlines. In the near future, better 
constraints will be available from the Baryon Acoustic Oscillation Sky Survey (BOSS \cite{boss}), 
part of SDSS-III. This will contain $160000$ quasar spectra between redshifts of $2.2$ and $3$, and
should further increase the statistical power of the Lyman-$\alpha$ forest. 
We plan to make forecasts for BOSS in forthcoming work \cite{soon}.

\section{Acknowledgements}
I thank my collaborators in this work, Hiranya Peiris, Matteo Viel and Licia Verde, for useful discussions. I am supported by STFC and by Pembroke College Cambridge. This work was performed using the Darwin Supercomputer of the University of Cambridge High Performance Computing Service (http://www.hpc.cam.ac.uk/), provided by Dell Inc. using Strategic Research Infrastructure Funding from the Higher Education Funding Council for England.

\end{document}